# Optimizing isotope and vacancy engineering in graphene ribbons to enhance the thermoelectric performance without degrading the electronic properties

Van-Truong Tran[1], Jérôme Saint-Martin[2], Philippe Dollfus[2] and Sebastian Volz[1]

[1]EM2C, CentraleSupélec, Université Paris Saclay, CNRS, 92295 Châtenay Malabry, France
[2]C2N, Université Paris-sud, Université Paris Saclay, CNRS, 91405 Orsay, France
van-truong.tran@ecp.fr and sebastian.volz@ecp.fr

**Abstract**

The enhancement of thermoelectric figure of merit *ZT* requires to either increase the power factor or reduce the phonon conductance, or even both. In graphene, the high phonon thermal conductivity is the main factor limiting the thermoelectric conversion. The common strategy to enhance *ZT* is therefore to introduce phonon scatterers to suppress the phonon conductance while retaining high electrical conductance and Seebeck coefficient. Although thermoelectric performance is eventually enhanced, all studies based on this strategy show a significant reduction of the electrical conductance, most often leading to a lower electronic performance. In this study we show that appropriate sources of disorder, including isotopes and vacancies at lowest electron density positions, can be used as phonon scatterers to reduce the phonon conductance in graphene ribbons without degrading the electrical conductance, particularly in the low-energy region which is the most important range for device operation. By means of atomistic calculations using semi-empirical Tight-Binding and Force Constant models in combination with Non-Equilibrium Green's function formalism, we show that the natural electronic properties of graphene ribbons can be fully preserved while their thermoelectric efficiency is strongly enhanced. For ribbons of width *M* = 5 dimer lines, room-temperature *ZT* is enhanced from less than 0.26 for defect-free ribbons to more than 2.5. This study is likely to set the milestones of a new generation of nano-devices with dual electronic/thermoelectric functionalities.

## Introduction

During the last decade, thermoelectricity has been the subject of renewed interest because of its expected contribution to current and future energy issues. [1,2] The conversion capacity of a thermoelectric material is reflected by the dimensionless figure of merit *ZT*, which was first introduced by Ioffe.[3–5] The best materials used to date in practice are compounds of Bismuth and Lead such as $Bi_2Te_3$, $Bi_2Se_3$, PbTe and their alloys, with a *ZT* ~ 1.[6] However, due to the slow advances in their efficiency, the toxicity of elements such as Se and Pb and also the limited resources in Tellurium,[7,8] the development of these materials in thermoelectricity remains limited.

In 1993, the studies of Hicks and Dresselhaus suggested that nanostructuring materials should offer better thermoelectric performance due to quantization effects.[9,10] Following these primary researches, a number of works have been carried out to target high thermoelectric performances in nanoscale-designed materials.[2,6,11] Among those studies, graphene and other 2D nano-materials have shown to be promising candidates for thermoelectric applications owing to their extraordinary electronic properties and especially the high flexibility they offer in tuning electronic and thermal properties, which can lead to large $ZT$, depending on material and structure design.[6,11]

In fact, 2D graphene has naturally poor thermoelectric properties because of its very high thermal conductivity and its gapless characteristic which induces extremely small Seebeck coefficients.[6,11–13] Interestingly, it has been shown that graphene is one of the most versatile materials in terms of thermoelectric properties that can be tuned from low to very high performance by nanostructuring.[6] First, graphene nano-ribbons (GNRs) can have higher $ZT$ compared to 2D sheets due to a bandgap opening induced by finite size effects, which results in larger Seebeck coefficients. The first investigation of thermoelectricity in ribbons was proposed by Ouyang and Guo in 2009.[14] They showed that $ZT$ of an armchair ribbon with 15 dimer lines along the width is enhanced compared to that of 2D graphene, though not exceeding 0.1. In a previous study, we pointed out that the maximum value $ZT = 0.35$ can be obtained for the narrowest armchair ribbon with a width of three dimer lines.[15]

To explore the thermoelectric properties of graphene with the aim of achieving higher $ZT$, different strategies based on more sophisticated structures than pristine nanoribbons have been proposed. For instance, mixed structures made of armchair and zigzag sections have been shown to exhibit $ZT_{max} \sim 1$ thanks to the mismatch of phonon modes and the resonant tunneling of electrons between the different sections.[16,17] In references [15,18], graphene or BN stubs (flakes) were attached into a graphene ribbon to generate interface phonon scattering. In the case of BN flakes attached to a GNR, $ZT = 0.81$ was reported for a ribbon of width $M = 5$ with even $ZT = 1.48$ in the presence of vacancies.[15] Vacancy or edge roughness disorder was also proved to be relevant to reduce the thermal conductance and achieve high thermoelectric performance.[19,20] Another interesting idea is to introduce nano-pores in the active region. In graphene ribbons with the presence of nano-pores $ZT$ can be dramatically enhanced up to 5 in some specific configurations.[21–23] Superlattice structures are also common designs to provide electronic transmissions with a step profile, which was proven to be associated with high $ZT$.[22,24,25] Chevron-type graphene ribbons including $^{14}C$ isotope doping were shown to have $ZT$ up to 3.25 at 800 K and about 2 at room temperature thanks to strongly reduced thermal conductance.[25] Graphene/h-Boron Nitride (G/BN) super-lattices have been also recently proposed to achieve high $ZT$ in straight ribbons.[7,26,27]

Although these strategies eventually yield high figures of merit $ZT$, we observe that a reduction of the phonon conductance is usually accompanied by a degradation of the electrical conductance which is likely to reduce the electronic performance of devices. Even in the case of Chevron-type ribbons with a fraction of $^{14}C$ isotopes,[25] mini-bandgaps are induced by the superlattice structure yielding a lower electrical conductance compared to that of the straight ribbon counterpart.

Since electronic properties of graphene are among its most intriguing and tunable ones,[28,29] it is strongly desirable to enhance its thermoelectric performance by decreasing its phonon conductance while keeping the electronic conduction properties unaltered.

In this article, on the basis of atomistic simulation, we show that by introducing appropriate disorders in armchair GNRs, very high thermoelectric performance can be achieved while fully preserving their natural excellent electronic properties. Indeed, the phonon conductance can be modulated strongly by introducing either $^{14}C$ isotopes in a $^{12}C$ lattice or vacancies. Isotopes reveal their effect mainly in the high-frequency region of the phonon spectrum while vacancies impact both the high and low-frequency regions. Additionally, if vacancy positions correspond to the sites of low electron density, the electronic conduction is shown to be unaffected. Combining all these effects, a maximum value of *ZT* can be tuned from 0.26 to values larger than 2.5 at room temperature for a ribbon width of $M = 5$ dimer lines.

## The modeling and methodology

We study armchair GNRs as sketched in Fig. 1. Fig. 1a presents a perfect (without disorder) structure made of carbon $^{12}C$ and each unit cell contains two chain lines $L_1$ and $L_2$. The active region includes $N_A$ unit cells and the length can be calculated as $L_A = 3a_0 \times N_A - a_0$ where $a_0 = 0.142$ nm is the distance between two nearest-neighbor atoms. The width of the ribbon is characterized by the number of dimer lines *M* along the *y* direction. Since each unit cell contains 2*M* atoms, the total number of atoms in the active region is $2M \times N_A$. Fig. 1b illustrates the cases in which the active region is either doped by $^{14}C$ isotopes or/and includes vacancies.

Although the first nearest neighbor Tight Binding (TB) calculation has been extensively used in many works to investigate electron properties, it has been shown that TB models involving up to the third nearest neighbor (3NN) interactions fit ab initio calculations more accurately. [30,31] Within this sophisticated calculation, the overlap matrix was also added, providing an optimized description of TB calculations in good agreement with ab initio results and suitable to large devices. [30,31]

In the present work, the 3NN TB model was employed for electron study using parameters taken from the work of Reich. [30] The Hamiltonian of the whole system can be written generally as

$$H\psi = ES\psi, \qquad (1)$$

where *H*, *S*, *E* and $\psi$ are the Hamiltonian, the overlap matrix, the eigen energy and the wave function, respectively. The matrix elements are calculated as follows: $H_{ii} = \varepsilon_i$ is the on-site energy at *i*th site, $S_{ii} = 1$ because the orbital wave function at the *i*th site is orthogonal itself, $H_{ij} = -t_{ij}$ and $S_{ij} = s_{ij}$ with $t_{ij}$ and $s_{ij}$ referring to the hoping and overlap parameters, respectively, between atoms at the *i*th and *j*th sites. Each couple of parameters $\{t_{ij}, s_{ij}\}$ must be selected as $\{t_0, s_0\}$, $\{t_1, s_1\}$ or $\{t_2, s_2\}$, i.e. as the first, second and third nearest-neighbor parameters, respectively, depending on the distance between atoms *i* and *j*.

For phonons, we employed a Force Constant (FC) model involving up to the four nearest neighbor interactions, which has provided a precise reproduction of the phonon dispersion of graphene obtained by ab initio methods and experimental measurements. [32] The force constant parameters were taken from Wirtz's work.[32] The motion equation of Newton's second law,[33] can be rewritten in a matrix form with the Dynamical matrix $D$, which can be seen as the matrix form of the Hamiltonian equation for phonons:[15,33]

$$DU = \omega^2 U, \qquad (2)$$

where $U$ is the matrix containing the vibrational amplitudes of all atoms and $\omega$ is the angular frequency. Since the coupling between two atoms $i$ and $j$ is described by a 3x3 tensor, the block elements of the dynamical matrix $D$ are finally [15]: $D^{ii}_{3\times3} = \sum_{j\neq i} \frac{K_{ij}}{M_i}$ and $D^{ij}_{3\times3} = -\frac{K_{ij}}{\sqrt{M_i M_j}}$ for $j \neq i$. The tensor coupling between atoms $i$ and $j$, $K_{ij}$ is defined from the force constant matrix parameters by a unitary in-plane rotation. [15,33]

When introducing substitutional $^{14}C$ isotopes in the $^{12}C$ lattice, the electron Hamiltonian was kept unchanged, while the dynamical matrix was adjusted by changing the mass at the positions of the $^{14}C$ isotopes.

To describe the presence of vacancies, we switched off all hoping or tensor couplings between a vacancy position and its neighbors. The onsite energy and mass at the vacancy positions were also set equal to infinity to ensure that the wave functions or vibrations vanish at the positions of vacancies.

The position of an isotope or a vacancy is defined by 3 indices ($n$, $L$, $m$) where $n$ is the position of the unit cell, with $n = 1:N_A$, $L$ is either $L_1$ or $L_2$, i.e. the line position in the unit cell, and $m$ is the position of the disorder site in the line $L_1$ (or $L_2$), i.e. $m = 1:M$.

As equations for electrons and phonons are similar (with $S = \mathbf{1}$ for the case of phonons), we employed in both cases the same method based on the Green's function formalism to compute the transport properties.[34] In practice, the Hamiltonian $H$ and the overlap matrix $S$ were divided into three parts $H_L$, $H_D$, $H_R$ and $S_L$, $S_D$, $S_R$ (similarly $D_L$, $D_D$, $D_R$ for phonons) as the Hamiltonians and overlap matrices of the left contact, device part and right contact, respectively. The coupling terms between the device and the two contacts are denoted $H_{DL}$, $H_{DR}$, $S_{DL}$, and $S_{DR}$ ($D_{DL}$, $D_{DR}$ for phonons). The Green's functions of the structure for electrons write:

$$G = \left[ E^+ . S_D - H_D - \Sigma^s_L - \Sigma^s_R \right]^{-1}, \qquad (3)$$

where $E^+ = E + i.\eta$ and $\eta$ refers to a positive infinitesimal number added to the energy to avoid the possible divergence of Green's functions and

$$\begin{aligned}\Sigma^s_L &= \left( E^+ . S_{DL} - H_{DL} \right) G^0_L \left( E^+ . S_{LD} - H_{LD} \right) \\ \Sigma^s_R &= \left( E^+ . S_{DR} - H_{DR} \right) G^0_R \left( E^+ . S_{RD} - H_{RD} \right)\end{aligned} \qquad (4)$$

define the surface self-energies contributed from the left and right contacts. $G^0_{L(R)}$ represents the surface Green's function of the isolated left (right) contact. Although Sancho's technique is widely used to compute the surface Green's functions,[35–37] the original method does not include the overlap matrices and a correction was thus applied to this method.[38,39]

For phonon calculation, a similar formalism was applied just by replacing energy $E$ by $\omega^2$, and $H_D$, $H_{DL}$, $H_{LD}$, $H_{DR}$, $H_{RD}$ by $D_D$, $D_{DL}$, $D_{LD}$, $D_{DR}$, $D_{RD}$, respectively. We also considered that $S_D = \mathbf{1}, S_{DL} = S_{LD} = S_{DR} = S_{RD} = \mathbf{0}$ for phonons.

The size of the device Green's function was reduced using the recursive technique.[36,37] Then electrons (phonons) transmission was computed as [15,34]

$$T_{e(p)} = Trace\left\{\Gamma^s_L\left[i\left(G_{11} - G_{11}^\dagger\right) - G_{11}\Gamma^s_L G_{11}^\dagger\right]\right\}, \qquad (5)$$

where $\Gamma^s_{L(R)} = i\left(\Sigma^s_{L(R)} - \Sigma^{s\dagger}_{L(R)}\right)$ denotes the surface injection rate at the left (right) contact. The electrical conductance, the Seebeck coefficient, the electron and phonon thermal conductance were computed using the Landauer-Onsager's approach [40], i.e.

$$\begin{aligned}
G_e(\mu, T) &= e^2 . L_0(\mu, T) \\
S(\mu, T) &= \frac{1}{e.T} \cdot \frac{L_1(\mu, T)}{L_0(\mu, T)} \\
\kappa_e(\mu, T) &= \frac{1}{T} \cdot \left[L_2(\mu, T) - \frac{L_1(\mu, T)^2}{L_0(\mu, T)}\right]
\end{aligned} \qquad (6)$$

Although the intermediate functions $L_n$ are usually formulated by making use of the Fermi function, we derived a more convenient form for practical use [15]

$$L_n(\mu, T) = \frac{1}{h} \int_{-\infty}^{+\infty} dE . T_e(E) . (2K_b T)^{n-1} . g^e_n(E, \mu, T), \qquad (7)$$

where $g^e_n(E, \mu, T) = \left(\frac{E - \mu}{2K_b T}\right)^n / \cosh^2\left(\frac{E - \mu}{2K_b T}\right)$ is a dimensionless function, which decays very quickly with respect to energy and can be used to estimate the boundaries of the integral (7). We also derived a similar form to calculate the phonon conductance

$$K_p = \frac{K_b}{2\pi} \int_0^\infty d\omega . T_p(\omega) . g^p(\omega, T), \qquad (8)$$

where $g^p(\omega,T) = \left(\dfrac{\hbar\omega}{2K_bT}\right)^2 / \sinh^2\left(\dfrac{\hbar\omega}{2K_bT}\right)$ and $K_b$ refers to the Boltzmann constant.

Once the electrical conductance, the Seebeck coefficient, the electron and phonon thermal conductances are obtained, the figure of merit $ZT$ is readily computed by the following equation [3,6,15]

$$ZT = \dfrac{G_e.S^2}{K_e + K_p}.T \qquad (9)$$

In some cases, thermoelectric ability of a material can be decomposed into fractional contributions of electrons and phonons separately as $ZT = ZT_e / (1 + K_p / K_e)$ with $ZT_e = (G_e.S^2 / K_e).T$ is considered as the electron figure of merit.

## Results and discussion

In this study, $^{14}$C isotopes were considered to appear randomly in the active region. We designed different configurations with distinct disorder positions but with the same density of disorder. Actually, for a given isotope doping density, ten different configurations were considered and eventually the average value for the transmission was taken as $\langle T_{e(p)} \rangle = \left(\sum_{i=1}^{10} T^i_{e(p)}\right)/10$. Then these averaged transmissions were used in equations (7) and (8). As shown later, the electronic properties remain unchanged by varying positions of scatterers along the device length, so that the notations of all electronic quantities were kept unchanged and without the average symbol $\langle\ \rangle$. The average phonon conductance is noted as $\langle K_p \rangle$.

### Isotopes: phonon scatterers transparent to electrons

Since $^{12}$C and $^{14}$C isotopes have the same electronic configuration and only differ in their mass, their random distribution generates phonon scattering without affecting the electronic properties. Thus the band structure, the electrical conductance and other electronic quantities are the same as those of pristine structures.

In Fig. 2, we display the room temperature electronic and thermoelectric properties of pristine GNRs. Fig. 2a shows the contributions of electrons to the different thermoelectric parameters, including the electrical conductance, the Seebeck coefficient and the electron thermal conductance for armchair ribbon of width $M = 5$. It is worth noting that this ribbon belongs to the group $M = 3p + 2$ which has been found to be semi-metallic (gapless) in the first nearest-neighbor (1NN) calculations.[41,42] Using third nearest-neighbor (3NN) calculation, it is noticeable from the inset that the transmission is equal to zero near the energy $E = 0$, which actually indicates the presence of a small energy gap in this structure. This result is in agreement with the predictions of *ab initio* calculations. [31] Thanks to the existence of a bandgap, the Seebeck coefficient is found to be significant and as high as 0.45 mV.K$^{-1}$, i.e. higher than the value of 80 µV.K$^{-1}$ in 2D graphene.[13]

In Fig. 2a, the power factor $P = G_e.S^2$ is plotted in the solid black line. The peak of the power factor is located at the crossing point of the Seebeck coefficient and the electrical conductance that is not the position of the peak of the Seebeck coefficient. The figures of merit *ZT* and *ZT$_e$* calculated by including or not, respectively, the lattice thermal conductance $K_p$ are displayed in Fig. 2b. Without the phonon contribution, the figure of merit *ZT$_e$* is much higher and reaches a maximum value of 12.93. However, the full *ZT* only reaches 0.26. The substantial difference between *ZT$_e$* and *ZT* reflects that the conductance $K_p$ contributes considerably to the total thermal conductance. Fig. 2a shows that $K_e$ at *ZT$_{max}$* is about 0.039 nW.K$^{-1}$ whereas $K_p$ can takes the value of 0.809 nW.K$^{-1}$ for *M* = 5, as shown in Fig. 2c. $K_p$ is thus about 20.7 times larger than the electron thermal conductance. Hence the phonon conductance is the predominant one and has to be reduced to enhance the thermoelectric performance.

In Fig. 2c, $K_p$ and *ZT$_{max}$* are also displayed for other ribbon widths. It can be observed that *ZT* reduces with increasing ribbon width since the phonon conductance increases. Narrow ribbons should therefore be considered for a higher *ZT*. Although thinner ribbons with *M* = 3 or 4 exhibit larger *ZT*, these ribbons are still difficult to synthesize in practice. The narrowest ribbon ever achieved had a thickness of *M* = 5 and was successfully synthesized by growing armchair GNRs in ultrahigh vacuum. [43]

To aim at the highest thermoelectric performance with a possible practical achievement, we hence focus on the case *M* = 5 in the course of the following discussion, unless otherwise stated.

To enhance *ZT*, $^{14}$C isotopes have been inserted as phonon dopants in the active region in order to generate phonon scattering and suppress $K_p$. To determine the most effective doping density to alter the phonon conductance, we plot the evolution of $\langle K_p \rangle$ as a function of the $^{14}$C doping percentage in Fig. 3. In Fig. 3a, it can be observed that $\langle K_p \rangle$ seems to reach a minimum for a density of about 70%. This apparent imbalance between $^{12}$C and $^{14}$C densities come from the fact that the leads are made of pure $^{12}$C graphene. To understand the effective range of frequency that isotopes affect, we plot the transmitted phonon spectrum in Fig. 3b. The spectrum of transmitted phonon energy is derived from the product of the transmission and the distribution function $g^p$: $T_p(\omega).g^p(\omega,T)$. As can be seen in Fig. 3b, all structures with different densities of $^{14}$C are significantly impacted in the high frequency range. The 0 to 350 cm$^{-1}$ frequency range remains weakly changed. For higher doping concentrations (dotted blue and dashed-dotted green lines), isotopes have an increasing influence in the low-frequency range and even completely suppress the vibrational modes of high frequency from 1200 cm$^{-1}$ to 1600 cm$^{-1}$ resulting in a decreased phonon conductance. The spectrally selective impact of isotopes on graphene phonon thermal conductance has not been previously highlighted so far. [44–46]

Now, we analyze the effect of the device length and temperature. In Fig. 4a we plot the phonon conductance $\langle K_p \rangle$ as a function of temperature for different device lengths ranging from $N_A$ = 30 ($L_A$ = 12.64 nm) to $N_A$ = 100 ($L_A$ = 42.46 nm), for an isotope density of 70%. Compared to the case of pristine pure $^{12}$C ribbon (black solid line), as expected and as already observed in other cases of disorder in GNRs,[19] increasing the device length in the presence of isotope disorder clearly tends to reduce the thermal conductance. For instance, at $T$ = 300 K, $\langle K_p \rangle$ is equal to 0.809 nW.K$^{-1}$ in the pristine structure but drops to 0.52 nW.K$^{-1}$ in a disordered device of length $L_A$ = 12.64 nm and even to 0.49 nW.K$^{-1}$ and 0.44 nW.K$^{-1}$ for devices of lengths 21.16 nm and 42.46 nm, respectively.

However, we clearly see also that at low temperature below 100 K the effect of isotope disorder is negligible whatever the device length. This behavior is consistent with the fact that at low temperature the width of the distribution function $g^p(\omega,T)$ covers only the low-frequency range (see inset of Fig. 4a), and with the previous results (Fig. 3b) showing that the transmission of such low-frequency modes is not impacted by isotope disorder.

To reveal the evolution of the phonon conductance with respect to the increase of the length and also exploit the potential of low phonon conductance in long devices, we consider with more care the length-dependence of $\langle K_p \rangle$ under the effect of disorder by introducing Matthiessen's rule which can be expressed in this context as[47]

$$\frac{1}{\tau} = \frac{1}{\tau_{lattice}} + \frac{1}{\tau_{disorders}}, \quad (10)$$

where $\tau$ is the total relaxation rate, $\tau_{lattice}$ and $\tau_{disorders}$ are the relaxation rate due to the lattice, i.e. phonon-phonon scattering, and to disorders, respectively. Within the ballistic regime, $\tau_{lattice} \approx \infty$ whereas in the presence of disorders $\tau_{disorders} \sim L_A / \Lambda$ where $\Lambda$ refers to the phonon mean free path. If we assume this mean free path to be independent of length, $1/\tau \sim 1/L_A$. It has been shown also that the phonon thermal conductivity $\sigma_p \sim \tau$,[47] and thus we have $1/\sigma_p \sim 1/L_A$. Since the thermal conductivity is proportional to $\langle K_p \rangle.L_A$, the quantity $1/(\langle K_p \rangle.L_A)$ is expected to be a linear function $1/L_A$. This quantity is plotted in Fig. 4b as a function of $1/L_A$ for different temperatures. The quasi-linear behavior is indeed observed and the results reveals that the conductivity at very long device length converges to a finite value independent on the temperature.

All these results strongly suggest that to further reduce the phonon conductance by degrading the phonon transmission in the low-frequency range, another scattering mechanism should be added to that induced by isotope disorder. We will see in the next section that vacancies appropriately positioned in the ribbon can be the appropriate scatterers in this respect.

**Introduction of 3$i$ vacancies**

Since isotopes cannot significantly suppress the phonon transmission at low frequencies, we investigate here the effect of vacancies for further reduction of the phonon conductance. We assume that vacancies can be controlled experimentally using focused electron beam techniques.[48,49] To minimize the effect of vacancies on the natural electronic performance, we first examine carefully the charge distribution over lattice sites.

It has been shown that the electronic properties of armchair GNRs vary for different groups of number of dimer lines $M$. We display in Fig. 5a the electron density at each site of lines $L_1$ and $L_2$ in a unit cell of pristine ribbons. The values $M = 12, 13, 14$ characterizes the three GNR width of groups $M = 3p$, $3p + 1$ and $3p + 2$, respectively, where $p$ is an integer number. From one edge to the center of a ribbon, the panels of Fig. 5a show that the charge at position $m = 3i$ is the lowest compared to other ones. For groups $3p$ and $3p + 1$, the ratio between highest and lowest charge is about 10 and 24, which means that the charge at positions $3i$ contributes to about 10% and 5%, respectively, to the total charge. Interestingly, this ratio is about $2.1 \times 10^3$ for the group $M = 3p + 2$. Our investigation reveals that in this group the charge at positions $3i$ only contributes about 0.05% of the total charge. This observation suggests that vacancies should be localized at position $m = 3i$ from the edges to minimize their effect on the electronic properties. More specifically, vacancies appearing at positions $3i$ in the structures $M = 3p + 2$ should not change at all the electronic properties due to extremely low charge at these lattice sites.

To verify the change of the electronic properties in the presence of vacancies, we have investigated all possible cases and summarized them in a picture displayed in Fig. 5b. Since positions $m = 3i+1$ and $3i+2$ have high charge localization, any vacancies at these positions will cause a dramatic change in the electrical conductance, e.g., a strong reduction of $G_e$ in the full range of energies as indicated in both panels of Fig. 5b. For the cases of vacancies at positions $m = 3i$, ribbons belonging to groups $M = 3p$ and $3p + 1$ still exhibit a smaller electrical conductance compared to that of the pristine structure as seen in the left panel of Fig. 5b.
Interestingly, ribbons of group $M = 3p + 2$ with vacancies at positions $m = 3i$ fully preserve their electrical conductance in low-energy region, which is the most relevant range for device operation. Due to the resulting unchanged electrical conductance, the Seebeck coefficient and the power factor, the thermoelectric performance also remains unaltered with the appearance of $3i$ vacancies.

To consider the effect of $3i$ vacancies on phonon transport, we only consider a 2% random vacancy density in a $3i$ line. Since each dimer line in the active region contains $2 \times N_A$ atoms, 2% $3i$ vacancies/line results in only $0.04 \times N_A$ $3i$ vacancies/line. Hence, for the structure having $N_A = 50$ unit cells (21.16 nm), only 2 vacancies were introduced. This small density of vacancies should thus not significantly distort the structure.

In Fig. 6a, the average phonon conductance of four structures is plotted as a function of temperature for the sake of comparison: the solid black line corresponds to the pristine structure, whereas the dashed, dotted and dashed-dotted lines refer to isotopes (70%) only, vacancies in line 3 (2%) only and to the combination of the two effects, respectively. Vacancies lead to a stronger reduction of the phonon conductance compared to the case of the sole isotope doping. Moreover, vacancies reveal their effect in the low-temperature range. This result is reflected by the phonon spectrum shown in Fig. 6b. The effect of vacancies is observed even at low frequencies (dotted blue line) in comparison to the case of isotope doping (dashed red line). Combining the two effects, the phonon conductance is further reduced (dash-dotted green line). The phonon spectra in Fig. 6b suggests that in the presence of vacancies, increasing the length of devices tends to suppress the transmission of phonon modes in the full range of frequencies. A very low phonon conductance is thus expected in long devices including both isotope disorder and vacancies.

To explore the combined effect of isotopes and vacancies in long devices, we have also plotted in Fig. 7a the inverse of the average phonon thermal conductivity versus the inverse of the device length. The numerical results were obtained for devices with length from $N_A = 50$ ($L_A = 21.16$ nm) to $N_A = 1400$ ($L_A = 596.26$ nm). The open symbol curves are numerical data while the solid lines are fitting curves. It can be seen that as in the case involving only isotope doping in Fig. 4b, we also observe a convergence here, however the numerical curves are not strictly linear. Actually, this non-linear behavior can be understood from the fact that the phonon mean free path $\Lambda$ may be not constant but length-dependent due to a competition between boundary scattering at side and axial boundaries. A change in length can cause a distortion in the total scattering and thus affecting $\Lambda$. The mean free path is almost constant when the device is long enough because of the dominant side scattering.

Since longer devices lead to lower thermal conductivity (and thus lower conductance), a higher thermoelectric performance is expected in long devices. To calculate $ZT$, we need to specify $\langle K_p \rangle$ in those long devices. The computation of Green's functions in long devices is very expensive so employing a fitting function is convenient to predict results in very long devices. We found that the fitting results are better with a quadratic fitting in the form of $1/(K_p L_A) = p_1 / L_A^2 + p_2 / L_A + p_3$. The quadratic term can be considered as a correction due to the length-dependence of the phonon mean free path and $p_3$ is obviously the inverse of the conductivity in infinitely-long devices. The set of fitting parameters $\{p_1, p_2, p_3\}$ for 100 K, 300 K, 400 K and 800 K are given in Table 2. We also applied a similar fitting for the results in Fig. 4b and the parameters are shown in Table 1. An analytical expression for the phonon conductance can subsequently be retrieved with $\langle K_p \rangle = L_A / (p_1 + p_2 . L_A + p_3 . L_A^2)$. The numerical (open symbols) and analytical (solid) results of the phonon conductance are plotted in Fig. 7b. The analytical results with the fitting parameters agree well with the numerical results. The analytical results and their extrapolation show that the phonon conductance is continually reduced and tends to saturate in very long devices.

**Enhancement of figure of merit *ZT***

In the presence of only 2% vacancies at positions 3*i*, all electronic properties of $M = 5$ ribbons remain unchanged, i.e. the same as in Fig. 2a. With the reduction of the phonon conductance due to both isotopes and vacancies as discussed above, the figure of merit is likely to be enhanced.

In Fig. 8a we show *ZT* at 300K as a function of the chemical energy $\mu$ for the device of length $L_A = 212.86$ nm for two cases (1) with 70% of isotopes and (2) with 70% of isotopes and 2% of additional 3*i* vacancies. With only isotope doping, *ZT* reaches a maximum value of 0.61 at either $\mu = -0.23$ eV or 0.03 eV. In the presence of additional vacancies, *ZT* is pushed strongly up to 1.52. The two peaks of *ZT* are now located at $\mu = -0.22$ eV and 0.02 eV. In these devices, the combined effect of both isotopes and vacancies thus leads to a *ZT* about 2.5 times larger than the one in the case of isotope disorder only.

In Fig. 8b, open symbols correspond to numerical results while solid lines refer to *ZT* resulting from the analytical formula of $\langle K_p \rangle$. Except for the cases of isotope doping only (open triangles and solid black lines), other results are shown for the cases when the two types of disorder are applied simultaneously. As observed from the black and green lines, the total effect is always larger than the effect of the sole isotope doping for a given length.

The additional effect of vacancies is also length-dependent, i.e. $ZT_{max}$ substantially increases with the increase of the device length $L_A$. The behavior at different temperatures is however different, i.e. the slope of the curves is larger when temperature varies from 100 K to 400 K. At 100 K, $ZT_{max}$ increases by 0.43 from 0.41 at $L_A = 21.16$ nm to 0.84 at $L_A = 298.06$ nm. In this range of lengths, at 300 K and 400 K, $ZT_{max}$ increases by 1.12 (from 0.66 to 1.78) and 1.25 (from 0.76 to 2.01), respectively. However the slope drops for $T = 800$ K since $ZT$ slowly increases from 0.48 for $L_A = 21.16$ nm to 0.56 for $L_A = 85.06$ nm and reaches 0.69 at 298.06 nm and then saturates. At $L_A = 596.3$ nm and $T = 400$ K, $ZT_{max}$ reaches a value of about 2.5 and the analytical results using the fitting form of $\langle K_p \rangle$ predict that $ZT_{max} \geq 3$ can be reached in longer devices if phonon-phonon scattering can still be neglected.

It is worth to note that $ZT_{max}$ at temperature $T = 800$ K is smaller than that at lower temperatures. This is a surprise since $\langle K_p \rangle$ is almost not dependent on temperature in very long devices as seen in Fig. 7b. It means that at higher temperatures, higher $ZT$ are expected because $ZT$ is proportional to $T$. Actually, this result can be understood if we pay attention to a quantity that is less important at lower temperatures in the total thermal conductance, which is the electron thermal conductance $K_e$.

In Fig. 9, we compare the electron and phonon conductances of the device of length $L_A = 596.26$ nm for a set of different temperatures. As can be seen, the electron thermal conductance $K_e$ (at $ZT_{max}$) increases exponentially with the increase of temperature, i.e., $K_e$ is about 0.0116 nW.K$^{-1}$ at 100 K, but at 300 K it contributes up to 0.0261 nW.K$^{-1}$ into the total thermal conductance. This value even jumps up to 0.503 nW.K$^{-1}$ at 800 K. Although $K_e$ at $ZT_{max}$ for $T < 800$ K is still smaller than $K_p$ of the pristine structures, it shows that in the structure with isotopes and vacancies, due to the dramatical reduction of $\langle K_p \rangle$, $K_e$ becomes larger than $\langle K_p \rangle$ at high temperatures, i.e., $K_e$ is predominant over $\langle K_p \rangle$ for $T \geq 450$ K. At 800K, $K_e$ is much larger than $\langle K_p \rangle$ and becomes the leading term therefore reducing the thermoelectric performance. These results suggest that the device performance are optimal in the medium range of temperature 300 to 400K, in which $K_e$ remains smaller than $\langle K_p \rangle$.

Since structures including isotope disorder and $3i$ vacancies yield a drastic drop of the phonon conductance, we try now to explore this effect for larger ribbon widths, which are more accessible in practice. In Fig. 10 we show results of $ZT_{max}$ for a ribbon of width $M = 11$ dimer lines. The open symbols and solid lines are the numerical results and the results using $\langle K_p \rangle$ fitting for the system of 50% doping $^{14}C$ and 5% vacancies $m = 6$. Since in the ribbon $M = 11$, the bandgap is smaller than in the ribbon $M = 5$, the device reaches the best performance at lower temperature, that is around 300 K instead of 400 K as observed in the case $M = 5$. For this device, $ZT_{max}$ reaches about 0.68 for the length of 170.3 nm at 300 K. $ZT > 1$ can be achieved for the devices of length $> 500$nm. Actually, $ZT > 1$ can be obtained even in short devices if we introduce more vacancies in other lines 3*i*. The filled triangular purple point is the result obtained for the device of length 170.3 nm when we have introduced 5% vacancies also in line *m* = 3 and *m* = 9 together with vacancies in line *m* = 6. The resulting figure of merit *ZT* is about 1.07 at room temperature and would be higher for longer devices.

## Conclusions

We have introduced a new concept yielding the enhancement of the thermoelectric figure of merit via thermal conductance reduction without degrading the electronic performance. We firstly showed that isotope disorder strongly suppress transmission of high frequency phonons. We then established a picture showing the effect of vacancy positions on the electronic properties. It came out that in position 3*i* in armchair nanoribbons of width *M* = 3*p* + 2, vacancies do not alter the electrical conductance and Seebeck coefficient while yielding a strong reduction in the transport of low frequency phonons.
The ribbons of width *M* = 5 in the presence of both isotope disorder and 3*i* vacancies have shown a dramatic improvement of figure of merit with a maximum value of *ZT* larger than 2.5 in long devices of lengths larger than 600 nm. Interestingly, these devices manifest their best performance in the medium temperature range from 300 K to 400 K.
We also demonstrated that larger ribbons of width *M* = 11 can achieve *ZT* > 1 thanks to isotope and vacancy engineering. These results can motivate the development of a new generation of nanodevices with dual thermoelectric/electronic functionalities.

## Acknowledgments

This work was supported by the TRANSFLEXTEG EU thermoelectric project.

Table 1: Fitting parameters for the inverse of the thermal conductivity in the case of isotope doping (Fig. 4b).

|  | 100 K | 300 K | 400 K | 800 K |
|---|---|---|---|---|
| $p_1$ (nm.K.nW$^{-1}$) | - 3.0391 | - 6.2710 | - 6.7673 | - 7.3256 |
| $p_2$ (K.nW$^{-1}$) | 3.9162 | 2.2290 | 2.0586 | 1.8518 |
| $p_3$ (K.nW$^{-1}$.nm$^{-1}$) | 0.0035 | 0.0047 | 0.0050 | 0.0054 |

Table 2: Fitting parameters for the inverse of the thermal conductivity in the presence of both isotopes and 3$i$ vacancies (Fig. 7).

|  | 100 K | 300 K | 400 K | 800 K |
|---|---|---|---|---|
| $p_1$ (nm.K.nW$^{-1}$) | -29.6450 | -28.3610 | -27.9020 | -27.2930 |
| $p_2$ (K.nW$^{-1}$) | 6.8577 | 4.2303 | 3.9148 | 3.5301 |
| $p_3$ (K.nW$^{-1}$.nm$^{-1}$) | 0.0204 | 0.0237 | 0.0242 | 0.0250 |

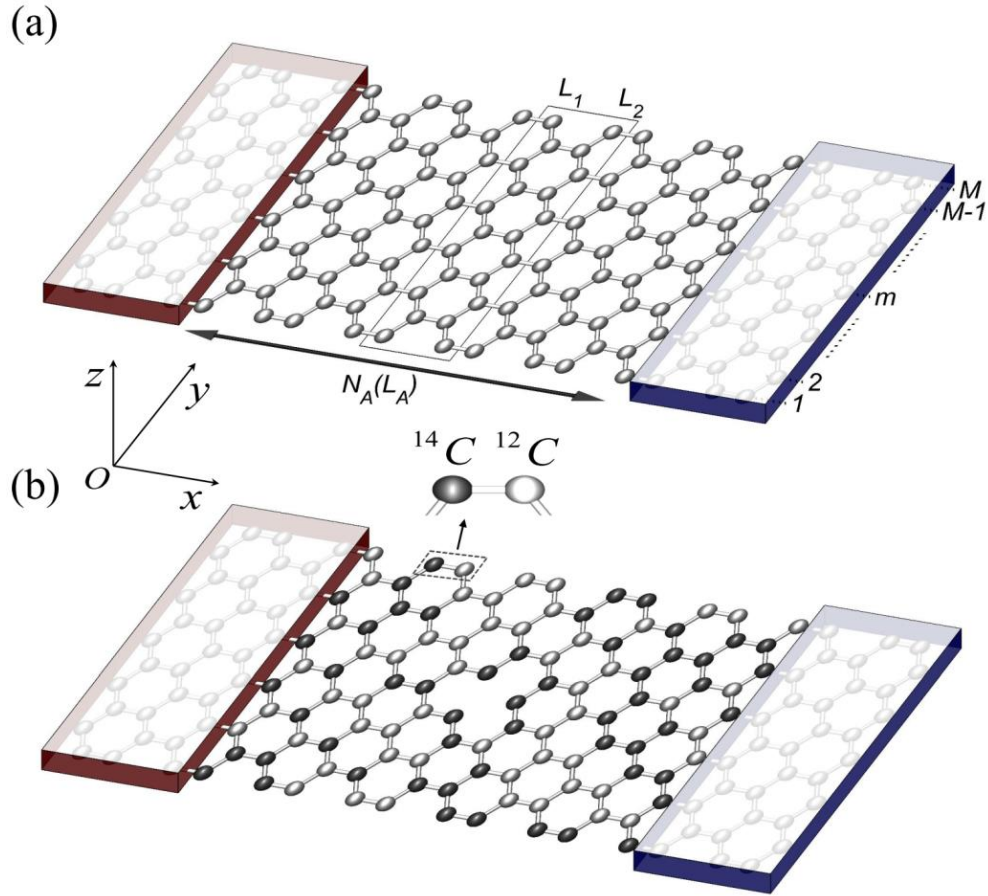

Fig. 1: Schematic view of devices made of armchair graphene nano-ribbons. (a) Pristine $^{12}$C structure without disorders, the width of ribbons is characterized by the number of dimer lines $M$, and the active region contains $N_A$ unit cells with the length $L_A$. (b) The active region includes a fraction of isotope $^{14}$C doping and a vacancy at position $m = 6$.

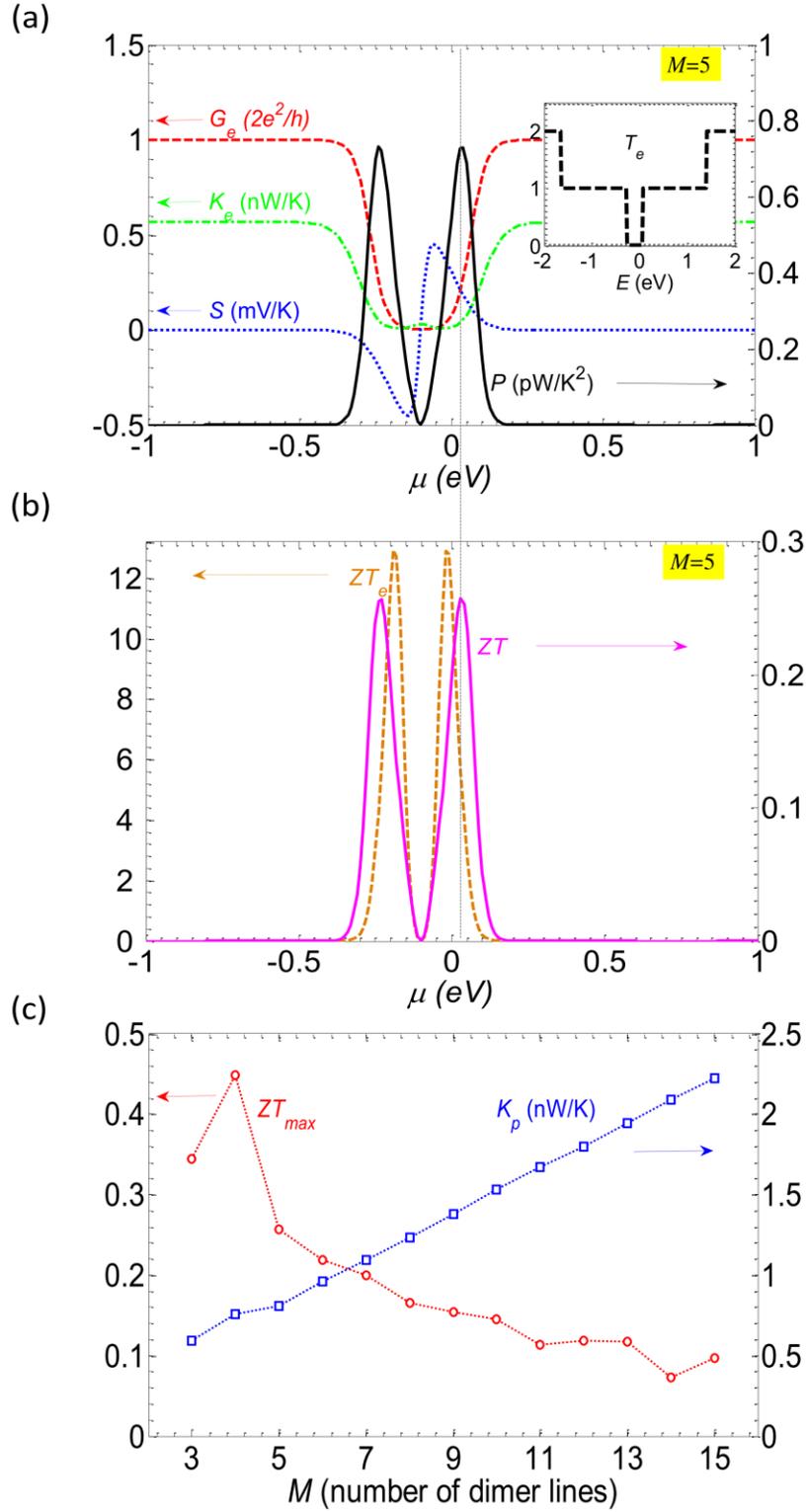

Fig. 2: Electronic and thermoelectric properties of a pristine structure of width $M = 5$ at room temperature. (a) The electrical conductance, the Seebeck coefficient, the electron thermal

conductance and the power factor plotted as functions of chemical potential. The inset reports the transmission spectrum. (b) The electronic and total figure of merit $ZT_e$ and $ZT$. (c) Maximum value of figure of merit $ZT_{max}$ and $K_p$ in different ribbons displayed as a function of $M$. In (a) and (b) the vertical dotted line indicates the position of $ZT_{max}$.

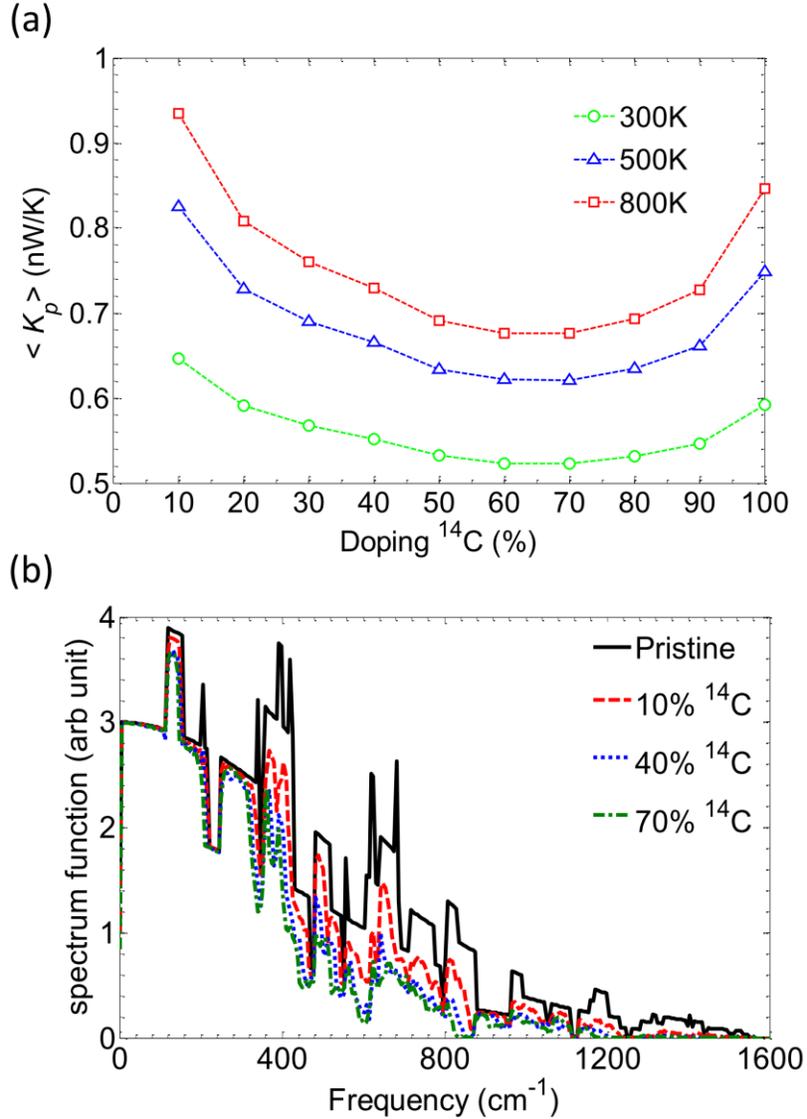

Fig. 3: (a) Average phonon conductance as a function of $^{14}C$ concentration. (b) Phonon energy spectrum versus frequency defining the frequency range impacted by isotopes. Data provided for $M = 5$, $N_A = 30$ ($L_A = 12.64$ nm) nanoribbons.

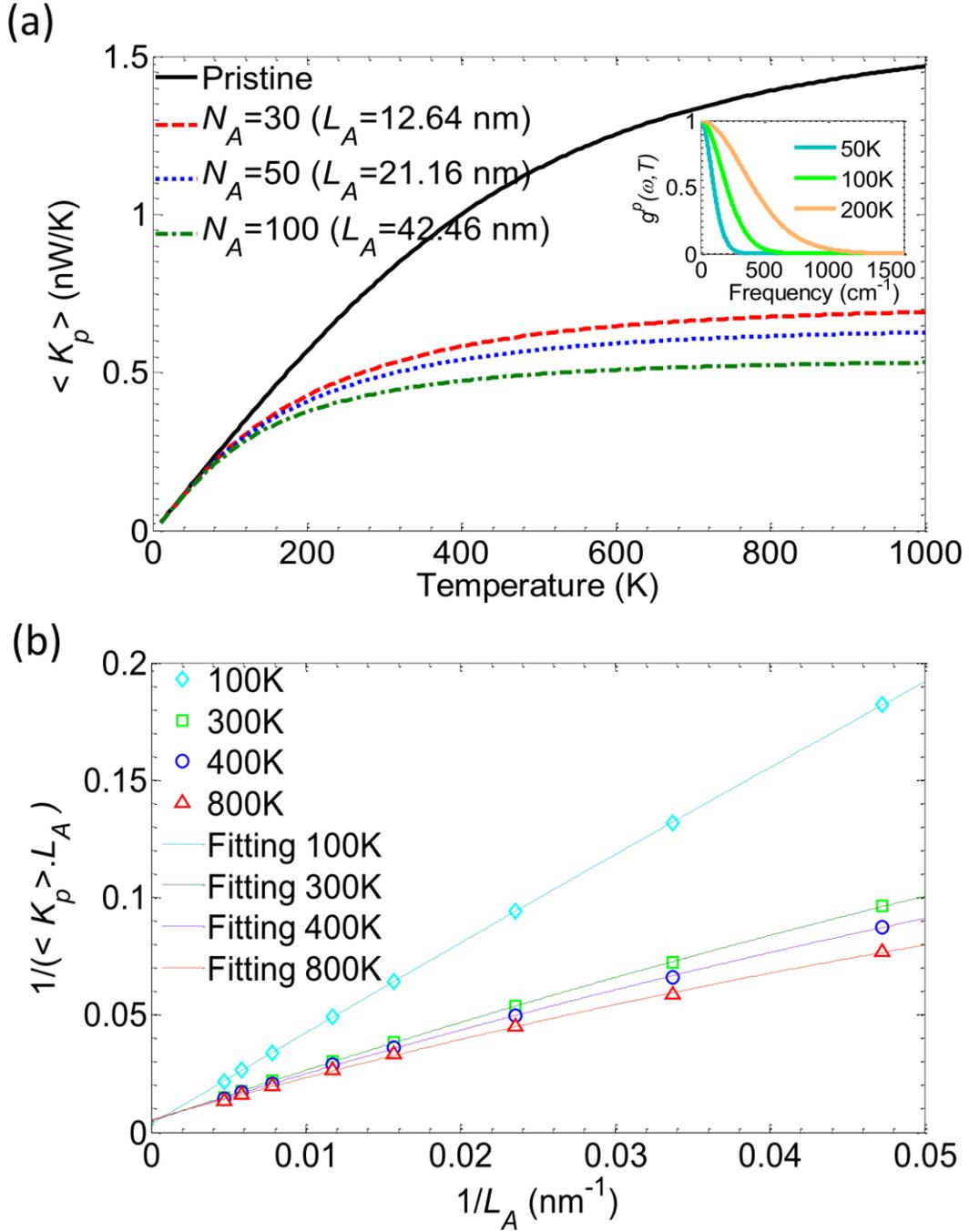

Fig. 4: (a) Average phonon conductance as a function of temperature for different device lengths with 70% isotope doping. The solid black corresponds to the pristine structure. The inset reports the distribution function $g^p(\omega, T)$ at different temperatures. (b) The inverse of the phonon thermal conductivity as a function of $1/L_A$. Symbols are numerical results and the solid lines are the fitting curves. Data provided for $M = 5$ nanoribbons.

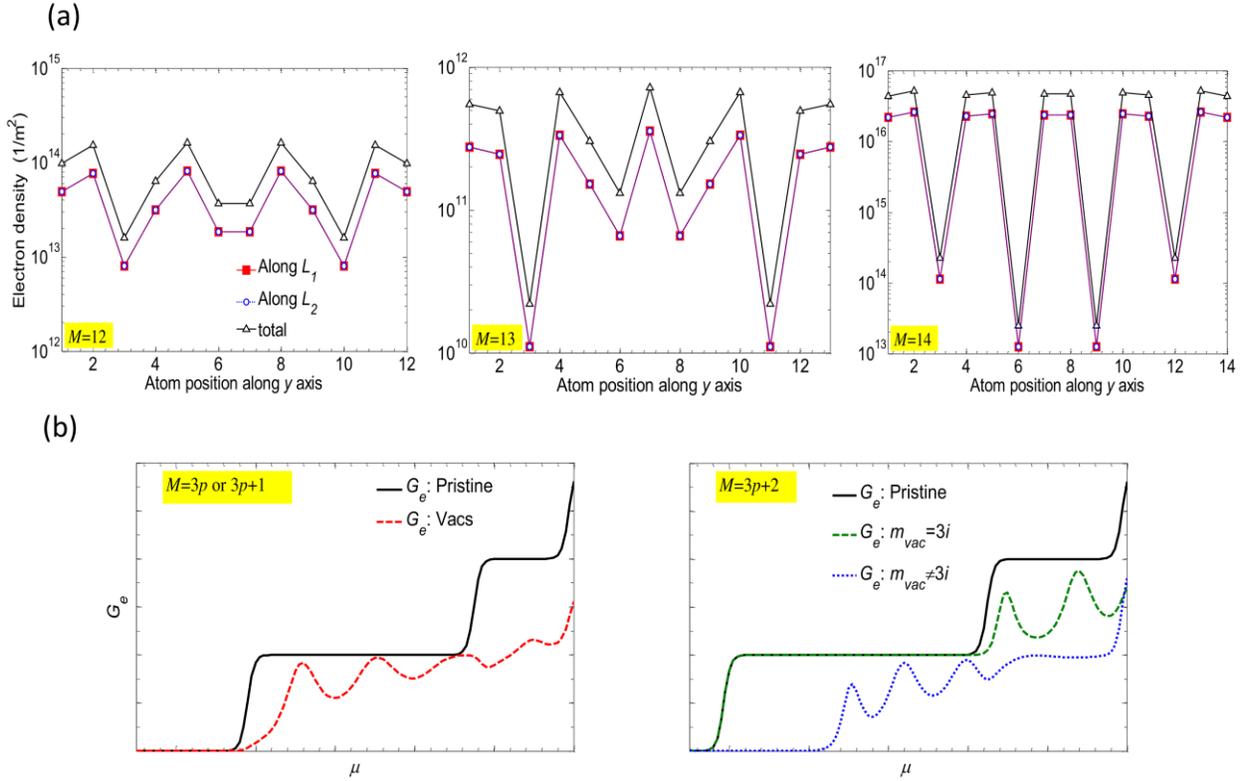

Fig. 5: (a) Spatial distribution of the electron density in pristine ribbons of width $M = 12$ (group $3p$), $M = 13$ (group $3p + 1$) and $M = 14$ (group $3p + 2$), charge is minimal at positions $m = 3i$ from one edge, especially in the group $3p + 2$. (b) Pictures of the effect of vacancies on the electrical conductance at different positions for 3 groups of ribbon widths.

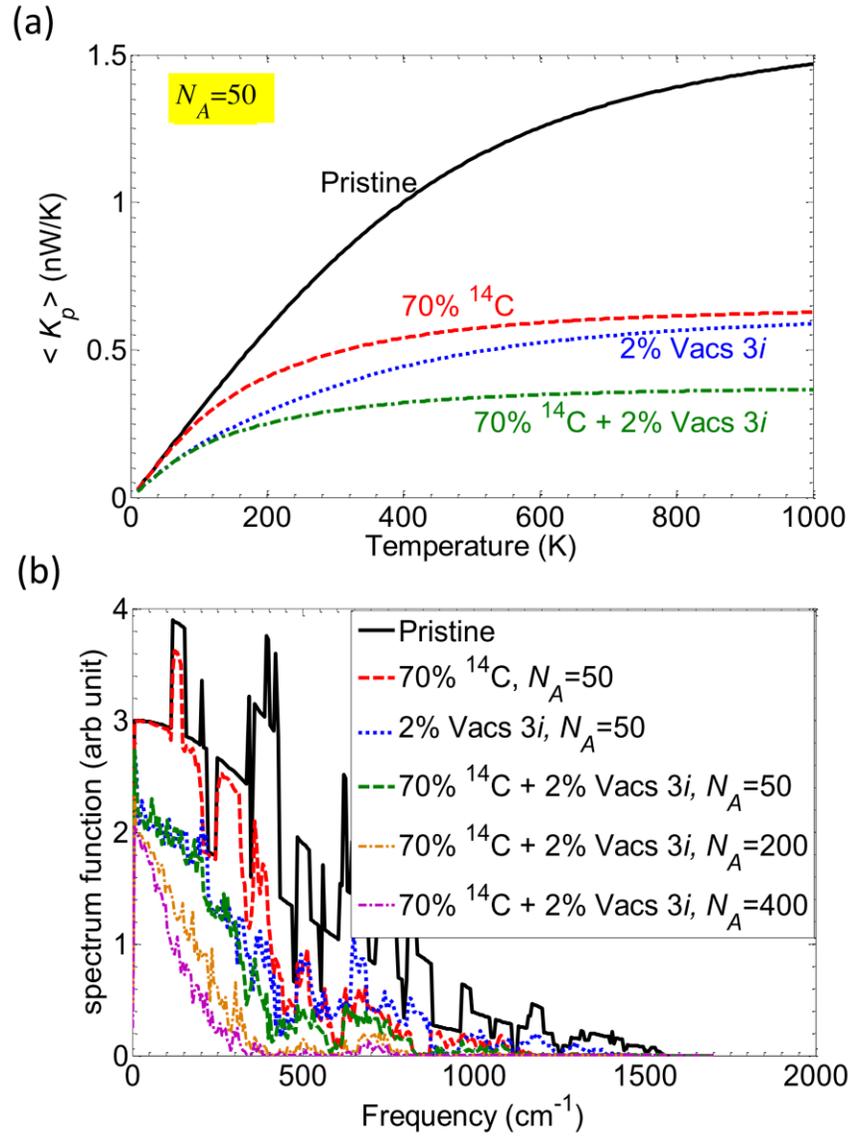

Fig. 6: (a) Average phonon conductance of four devices: pristine (solid black), with 70% isotope only (dashed red), 2% $3i$ vacancies only (dotted blue) and for both types of disorder (dashed-dotted green). (b) Spectrum functions of different disordered structures. Data provided for $M = 5$ nanoribbons.

.

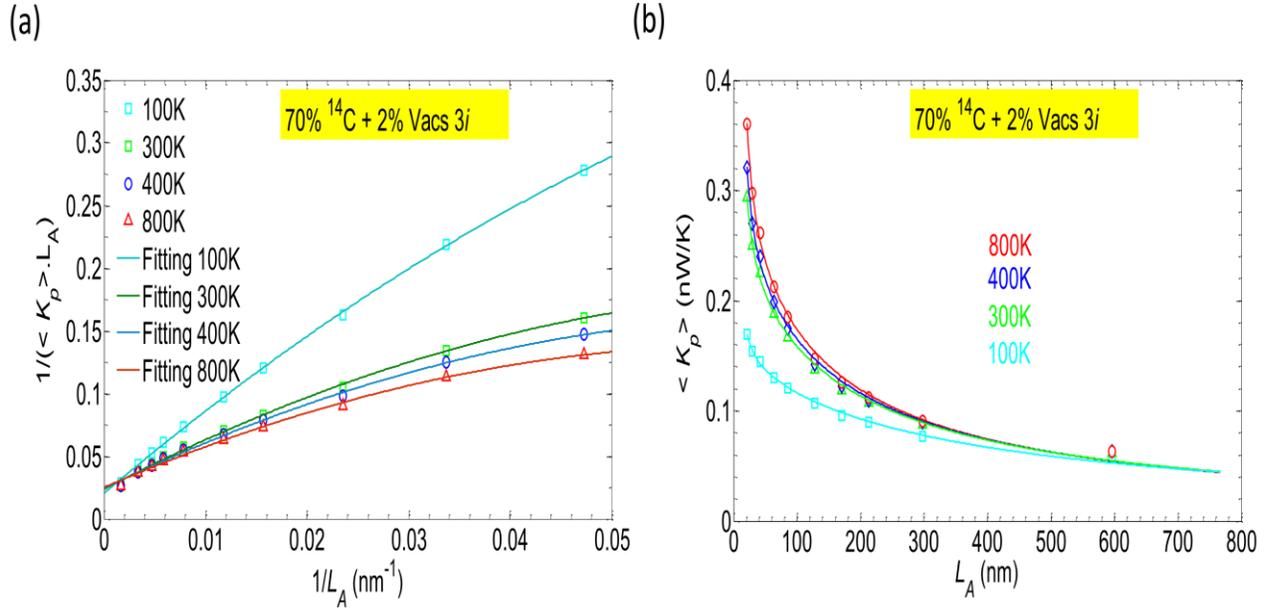

Fig. 7: (a) The inverse of the thermal conductivity as a function of $1/L_A$. (b) Average phonon conductance as a function of the active length $L_A$. In the both panels, symbols are numerical results and the solid lines are fitting/analytical results. In figure (b), the filled symbols are additional numerical results for comparison with results predicted by solid lines. Data provided for $M = 5$ nanoribbons.

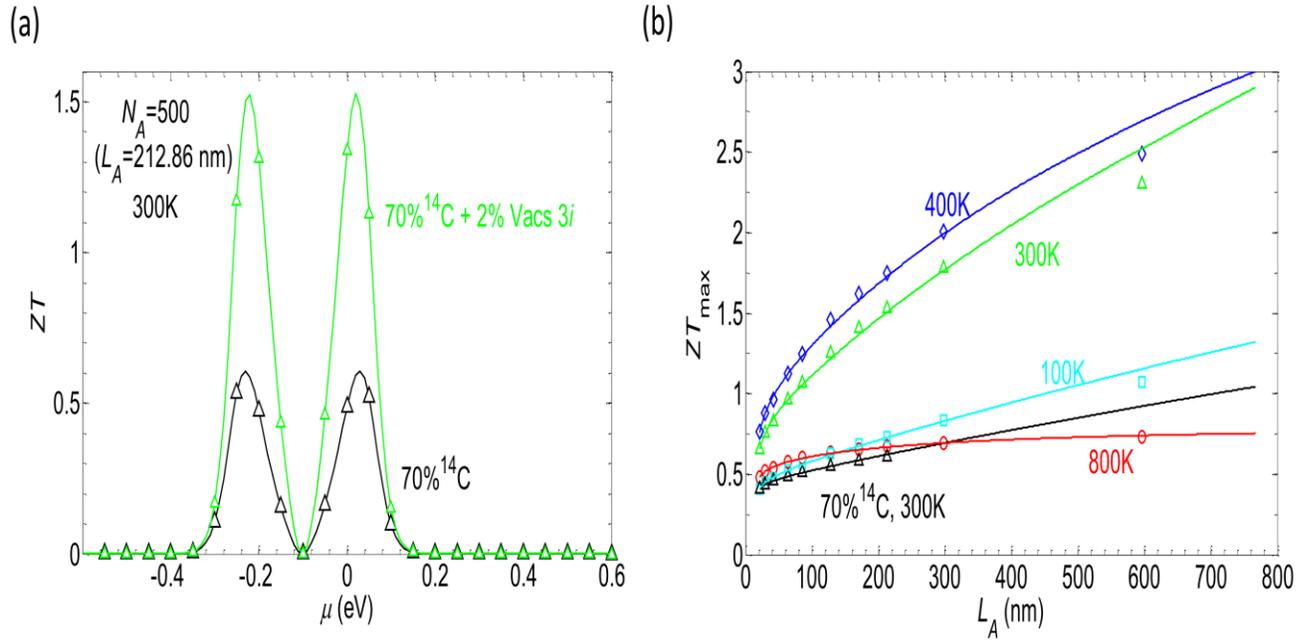

Fig. 8: (a) Figure of merit *ZT* as a function of chemical energy demonstrated for an isotope doped device of length 212.86 nm with and without additional vacancies 3*i*. (b) $ZT_{max}$ as a function of the device length, except for the black lines corresponding to isotopes only, other curves refer to of the combined effects of 70% isotope doping plus 2% 3*i* vacancies, the symbols are numerical data and the solid lines are the fitting curves. Data provided for $M = 5$ nanoribbons.

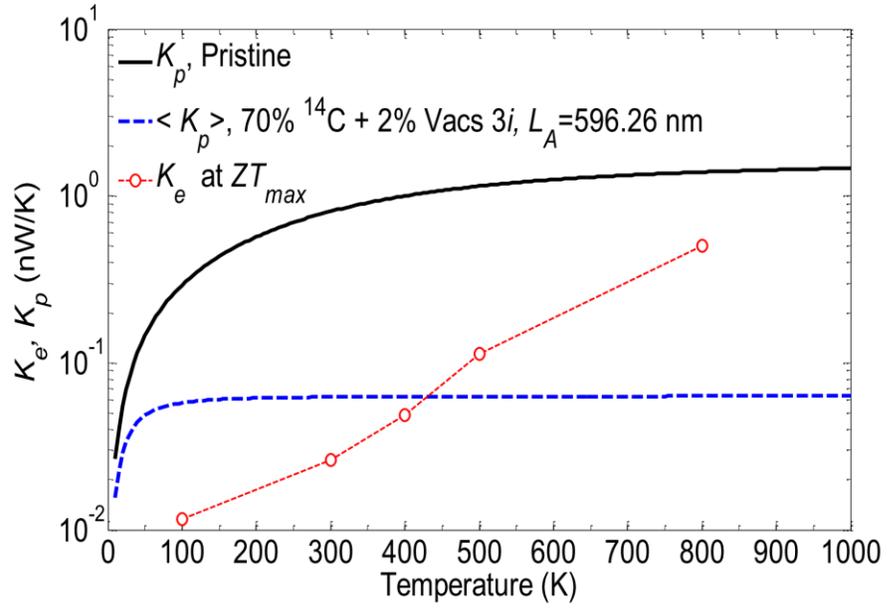

Fig. 9: Comparison between the phonon and electron thermal conductances at positions of $ZT_{max}$ for different temperatures. Data provided for $M = 5$ and $N_A = 1400$ ($L_A = 596.26$ nm) nanoribbons.

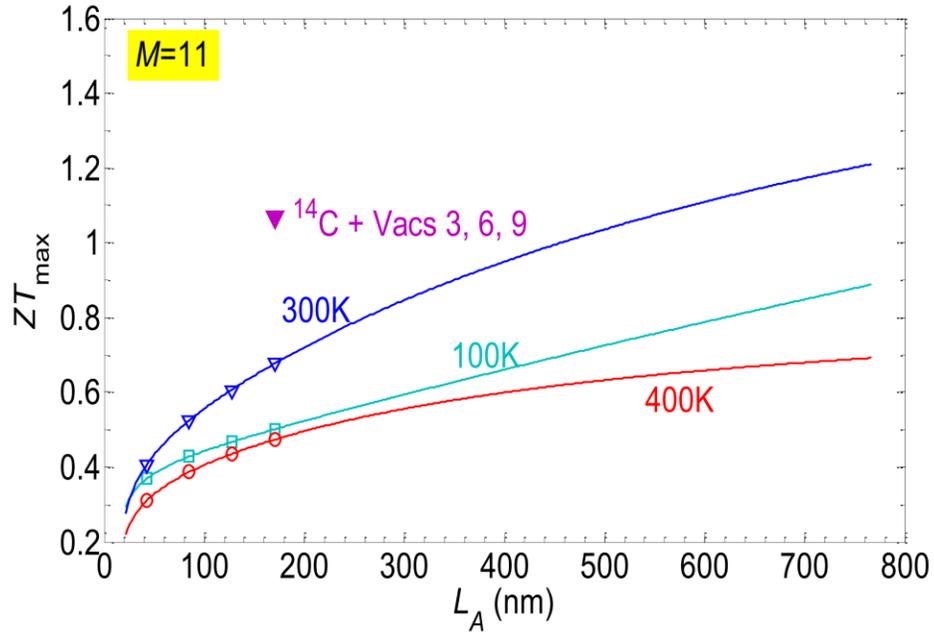

Fig. 10: $ZT_{max}$ as a function of the device length for the case of a larger ribbon with $M = 11$. The open symbols and solid lines are numerical data and fitting curves for the case of 50% isotope doping and 5% vacancies in the $m = 6$ line. The filled symbol is the numerical results of the device with 50% isotope doping and 5% vacancies in lines $m = 3, 6, 9$.